\documentclass[aps,prb,twocolumn,superscriptaddress,showpacs]{revtex4}

\usepackage{graphicx}
\usepackage{amsmath,amsfonts,amssymb}

\newcommand{\Ef}{{$E_{\text{F}}$}}

\begin{document}

\title{Quantum modulation of the Kondo resonance of Co adatoms on Cu/Co/Cu(100)}

\author{Takashi Uchihashi}
\email{UCHIHASHI.Takashi@nims.go.jp}
\affiliation{Institut f\"ur Experimentelle und Angewandte Physik, Christian-Albrechts-Universit\"at zu Kiel, D-24098 Kiel, Germany}
\affiliation{Nano System Functionality Center, National Institute for Materials Science, Tsukuba 305-0044, Japan}
\author{Jianwei Zhang}
\affiliation{Institut f\"ur Experimentelle und Angewandte Physik, Christian-Albrechts-Universit\"at zu Kiel, D-24098 Kiel, Germany}
\author{J\"org Kr\"oger}
\affiliation{Institut f\"ur Experimentelle und Angewandte Physik, Christian-Albrechts-Universit\"at zu Kiel, D-24098 Kiel, Germany}
\author{Richard Berndt}
\affiliation{Institut f\"ur Experimentelle und Angewandte Physik, Christian-Albrechts-Universit\"at zu Kiel, D-24098 Kiel, Germany}

\date{\today}

\begin{abstract}
Low-temperature scanning tunneling spectroscopy  reveals that the Kondo
temperature $T_{\text{K}}$ of Co atoms adsorbed on Cu/Co/Cu(100) multilayers
varies between $60\,{\text{K}}$ and  $134\,\text{K}$ as the Cu film thickness
decreases from $20$ to $5$ atomic layers. 
The observed change of $T_{\text{K}}$  is attributed to a
variation of the density of states at the Fermi level $\rho_{\text{F}}$ induced by quantum
well states confined to the Cu film.  A model calculation based on the quantum
oscillations of $\rho_{\text{F}}$ at the belly and the neck of the
Cu Fermi surface reproduces most of the features in the measured variation of $T_{\text{K}}$.
\end{abstract}


\maketitle

Magnetic atoms interacting with their environment are fundamentally important
in extensive fields of modern physics. Among all, the Kondo effect, a
representative phenomenon resulting from the exchange interaction of a local
spin with surrounding conduction electrons, is observed below a characteristic Kondo temperature $T_{\text{K}}$. \cite{HewsonYoshida_Kondo} One of its
hallmarks is the Kondo resonance whose width is given by
$\text{k}_{\text{B}}T_{\text{K}}$ ($\text{k}_{\text{B}}$: Boltzmann's constant).
The Kondo resonance may be directly investigated using low-temperature scanning
tunneling microscopy (STM) \cite{Li_KondoCeAg,Madhavan_KondoCoAu,Manoharan_QuantumMirage}
or various quantum dots.
\cite{Wiel_KondoQD,Nygard_KondoCNT,ParkLiang_KondoMolecule} Furthermore,
artificial modification of the Kondo effect was investigated using
atom manipulation \cite{Kliewer_Newns,Limot_Step,Chen_KondoDimer} and controlled
tip contact. \cite{Neel_KondoContact} More recently, a molecular Kondo effect
has been investigated. \cite{Hou_KondoMolecule,Iancu_KondoMolecule,Wahl_KondoExchange,Xue_KondoPbQW}
Apart from the Kondo effect, the recent advancement of the spintronics has raised a strong interest in spin manipulation and in tailoring exchange interaction at atomic scale. \cite{Meier_ExchangeSPSTM,Brovko_ExchangeTheory}
The observation of spin-related phenomena using low temperature STM and their control with artificial nanostructures are crucial for realizing such promises.

Here we use a model system for tuning the single-adatom Kondo temperature by means of variations of the local density of states at the Fermi level.
In particular, single Co atoms are
adsorbed on multilayers of Cu and Co grown on a Cu(100) surface and modifications
of the Kondo resonance are measured as a function of the Cu layer thickness.
Scanning tunneling spectroscopy is used to identify quantum well (QW) states
of the Cu film. The absolute thickness of Cu overlayers,
$d_{\text{Cu}}$, is locally determined from characteristic QW state energies.
The Kondo temperature determined from spectra of the differential
conductance ($\text{d}I/\text{d}V$) taken on single Co adatoms varies between
$60\,\text{K}$ and $134\,\text{K}$ with decreasing Cu  coverage
from $20$ to $5$ atomic layers. 
Much of this variation is reproduced by a model calculation based on the quantum oscillations at the belly and the neck of the Cu Fermi surface. 
The phases of these oscillations are found to deviate from a theoretical prediction for Cu/Co/Cu(100) multilayers, suggesting the effect of Co adatoms on the phase shift at the Cu/vacuum interface.

The experiments were performed with a home-built low temperature STM operated
at $7\,{\text{K}}$ and in  ultrahigh vacuum with a base pressure of
$10^{-9}\,{\text{Pa}}$. Samples and W tips
were prepared by annealing and argon ion bombardment. Crystalline order was
checked by low-energy electron diffraction (LEED) while preliminary estimates
of Co and Cu film thicknesses were performed by Auger electron spectroscopy (AES).
Spectra of $\text{d}I/\text{d}V$ were acquired by standard lock-in detection.
The clean Cu(100) surface was covered at
room temperature (RT) with $10\,\text{ML}$ of Co using an electron beam
evaporator and an evaporant of $99.99\,\%$
purity. We define a monolayer (ML) as one Co atom per Cu atom. Subsequently,
Cu was deposited at RT on the Co surface from a copper wire of $99.995\,\%$
purity wrapped around a tungsten filament. High deposition rates of
$\approx 6\,\text{ML}\,\text{min}^{-1}$ were necessary to suppress heating
from the filament and concomitant intermixing of Cu and Co. \cite{CoSegregation}
Co atoms were deposited on the cold surface by electron
beam evaporation.
\begin{figure}
  \includegraphics[width=80mm]{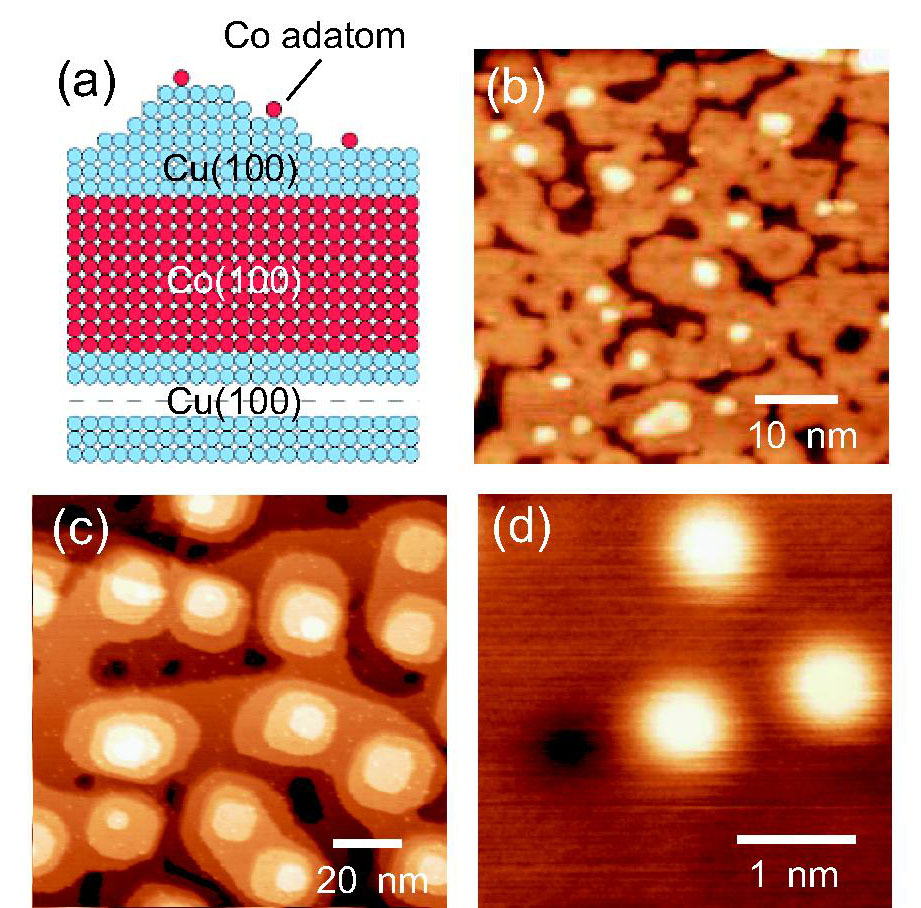}
  \caption{(Color online) (a) Sketch of Cu/Co/Cu(100) multilayer with
  single adsorbed Co atoms. STM images of (b) a 10-ML-thick Co overlayer grown
  on a Cu(100) substrate ($I=100\,\text{pA}$, $V=200\,\text{mV}$), (c) a
  15-ML-thick Cu overlayer grown on a Co/Cu(100) multilayer ($I=100\,\text{pA}$,
  $V=200\,\text{mV}$), and (d) of Co adatoms deposited on a Cu/Co/Cu(100)
  multilayer ($I=100\,\text{pA}$, $V=300\,\text{mV}$).}
  \label{Fig1}
\end{figure}

We adopted a Cu/Co/Cu(100) system as a substrate for tuning the Kondo effect
because extensive studies on this multilayer system are available. \cite{Qiu_PESQW}
An excellent lattice matching enables
epitaxial growth of a face-centered cubic Co(100) layer on Cu(100), rather
than a bulk-like hexagonal close-packed Co layer, and a subsequent epitaxial
growth of Cu(100) on Co(100) (Fig.\,\ref{Fig1}(a)).
Figure \ref{Fig1}(b) presents a constant-current STM image of
Cu(100) with a Co coverage of $10\,\text{ML}$ which reveals layer-by-layer
epitaxial growth of Co. A subsequently deposited Cu layer exhibits
pyramid-shaped islands with a lateral size of $\approx 20\,\text{nm}$
(Fig.\,\ref{Fig1}(c)). The resulting layered surface is an ideal
platform for acquiring spectra at a variety of Cu layer thicknesses without
changing the image area. Figure \ref{Fig1}(d) is a typical STM image
of  single Co adatoms on a Cu/Co/Cu(100) multilayer. Although the density
of adatoms is rather high (typically $0.2\,\text{atoms}/\text{nm}^2$), the Kondo effect remains intact as far as single
adatoms appear separated from each other in an STM image. \cite{Chen_KondoDimer}

\begin{figure}
  \includegraphics[width=80mm]{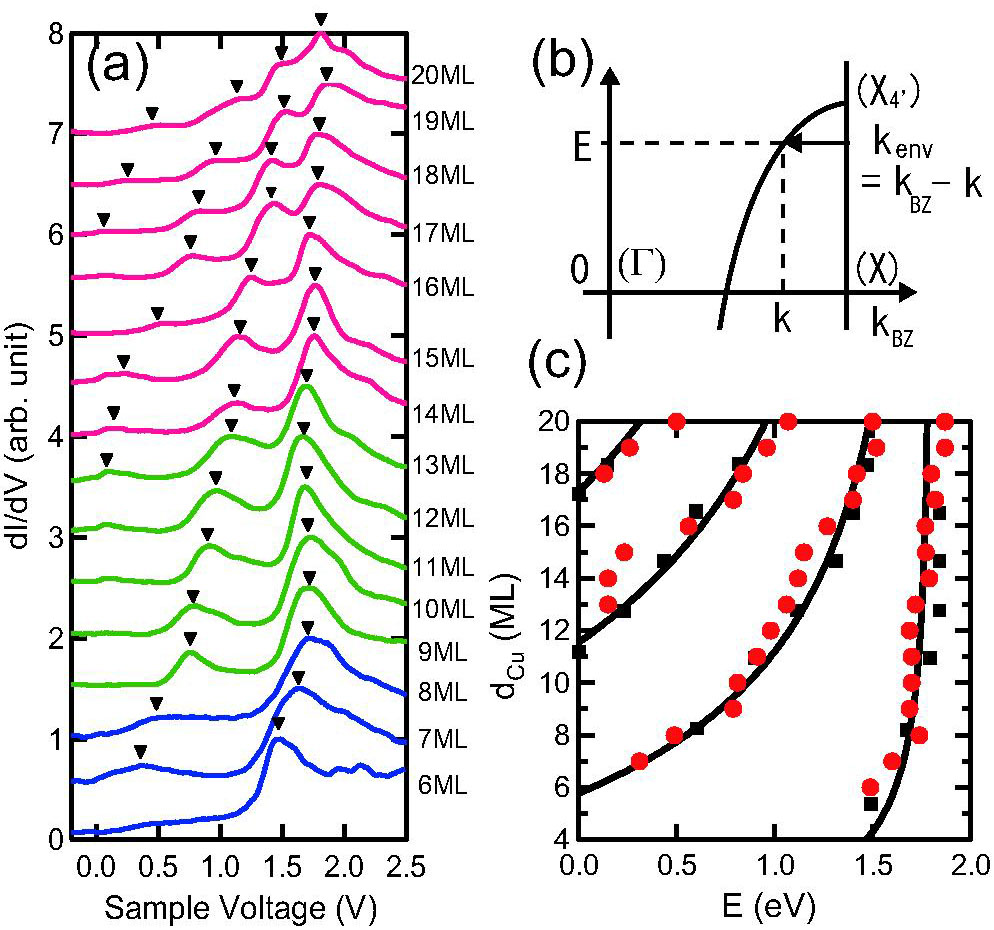}
  \caption{(Color online) (a) Series of $\text{d}I/\text{d}V$ spectra
  measured on different surfaces of Cu/Co/Cu(100) multilayers. Prior to
  spectroscopy the tunneling gap was set at $1\,\text{nA}$ and $2.5\,\text{V}$.
  Spectra are offset for clarity. The labels beside the graph show the
  thicknesses of the Cu overlayers determined by referring to theoretical
  curves in (c). The relative thicknesses within each group
  ($6$ -- $8\,\text{ML}$, $9$ -- $13\,\text{ML}$, $14$ -- $20\,\text{ML}$)
  were determined by a simultaneously observed surface topography. (b) Schematic
  dispersion curve of the Cu $sp$ band along the $[100]$ direction
  ($\overline{\Gamma}\overline{X}$). (c) Peak energies of $\text{d}I/\text{d}V$
  spectra in Fig.\,\ref{Fig2}(a) as a function of Cu overlayer thickness
  $d_{\text{Cu}}$ (dots). Calculated energies of QW states (solid lines)
  and IPES data (squares) \cite{Ortega_PESQW} are included.}
  \label{Fig2}
\end{figure}

While a nominal thickness of a Cu overlayer can be estimated from
the deposited amount of Cu, its local thickness varies substantially.
We determined the local thickness from the energies of QW states
of the Cu overlayer.
\cite{Qiu_PESQW,Ortega_PESQW,Bruno_ExchangeTheoryQW,Crampin_CoCuQW,Kawakami_PESQW}
Since the Cu Fermi surface is located near the Brillouin zone boundary, the
wave function of an electronic state near the Fermi level \Ef\
with wave number of $k$  is modulated by
an envelope function with
$k_{\text{env}}=k_{\text{BZ}} - k$, where $k_{\text{BZ}}$ is the wave number
at the Brillouin zone boundary. \cite{Ortega_PESQW,Kawakami_PESQW} QW states
occur when electrons are reflected at the two interfaces and interfere
constructively.
For an electron state with $k_{\text{env}}$ and energy $E$, this condition
is satisfied when the layer thickness $d_n$ is given by \cite{Ortega_PESQW}
\begin{equation}
  d_n=\left[ n-1+\frac{\Phi(E)}{2\pi} \right]\frac{k_{\text{BZ}}}{k_{\text{env}}},
  \label{eq:k-d_relation}
\end{equation}
where $\Phi(E)$ is the total phase shift caused by the reflections and $n$ is
an integer. For $E$, $k_{\text{env}}$ is determined by the energy
dispersion of the relevant $sp$ band (see Fig.\,\ref{Fig2}(b)). Because the dispersion
of the Cu $sp$ band is precisely known near \Ef, the thickness of the Cu layer
can be determined from the energies of the QW states.

Figure \ref{Fig2}(a) shows a series of $\text{d}I/\text{d}V$ spectra taken on
different surfaces of Cu/Co/Cu(100) multilayers. Clear peaks are visible within
the range of $0$ to $2.0\,\text{V}$ indicating the
presence of QW states in this energy range. \cite{footnote_QWpeaks} The local
thickness of the Cu overlayer was determined
by comparison of the experimentally obtained peak energies with
calculations. The energy dispersion of the Cu $sp$ band along the $[100]$
direction near \Ef\ can be described by the two-band nearly-free electron model:
\cite{Qiu_PESQW,Kawakami_PESQW}
\begin{equation}
  \frac{k_{\text{env}}}{k_{\text{BZ}}}=\sqrt{1+\frac{E+E_{\text{F}}}{G} - \sqrt{\frac{4(E+E_{\text{F}})}{G} + \left(\frac{U}{G}\right)^2}},
  \label{eq:energy_dispersion}
\end{equation}
where $E$ is the QW state energy measured relative to the Fermi level,
$G=\hbar^2 k_{\text{BZ}}^2/2m^{*}$ ($m^{*}$ is the effective mass of the
electron), $U$ is half the energy gap at the Brillouin zone boundary. Following
the analysis of QW states observed with inverse photoemission spectroscopy
(IPES) by Ortega \textit{et al.}, \cite{Ortega_PESQW}
we insert $E_{\text{F}}=7.39\,\text{eV}$, $G=12.27\,\text{eV}$, and
$U=3.08\,\text{eV}$ into Eq.\,(\ref{eq:energy_dispersion}) and a linear
phase-energy relation $\Phi(E) = 0.35\pi\,\text{eV}^{-1}\,\times\,E$ into Eq.\,(\ref{eq:k-d_relation}).
Figure \ref{Fig2}(c) displays the calculated thickness of a Cu overlayer as a function
of QW state energies (solid lines) together with IPES data (squares).
\cite{Ortega_PESQW} An absolute thickness was determined for each spectrum
in Fig.\,\ref{Fig2}(a) by referring to these theoretical curves (see the thicknesses
labeled beside the graph). Note that the \textit{relative} thicknesses within
a group ($6$ -- $8\,\text{ML}$, $9$ -- $13\,\text{ML}$, $14$ -- $20\,\text{ML}$)
were determined by a simultaneously observed surface topography; only an
identical offset was added to relative thicknesses within each group to fit
\textit{all} peak energies of \textit{all} spectra to the theoretical
curves. The results are summarized in Fig.\,\ref{Fig2}(c) (dots). The excellent
agreement between the experiment and the theory shows the validity of the present
analysis. The thickness determination error is estimated to $\pm 1\,\text{ML}$.

After determining $d_{\text{Cu}}$, we measured $\text{d}I/\text{d}V$
spectra on single Co atoms on Cu/Co/Cu(100) multilayers.
Before
performing spectroscopy, the STM tip was shaped by touching a Cu surface area
until Co adatoms appeared circular and $\text{d}I/\text{d}V$ spectra of a bare
Cu surface became featureless around zero sample voltage. For analysis of the
Kondo resonance, remaining artifacts due to the tip electronic structure were
removed by dividing spectra of single Co atoms by an averaged spectrum
of the clean Cu surface on the same terrace.
\begin{figure}
  \includegraphics[width=80mm]{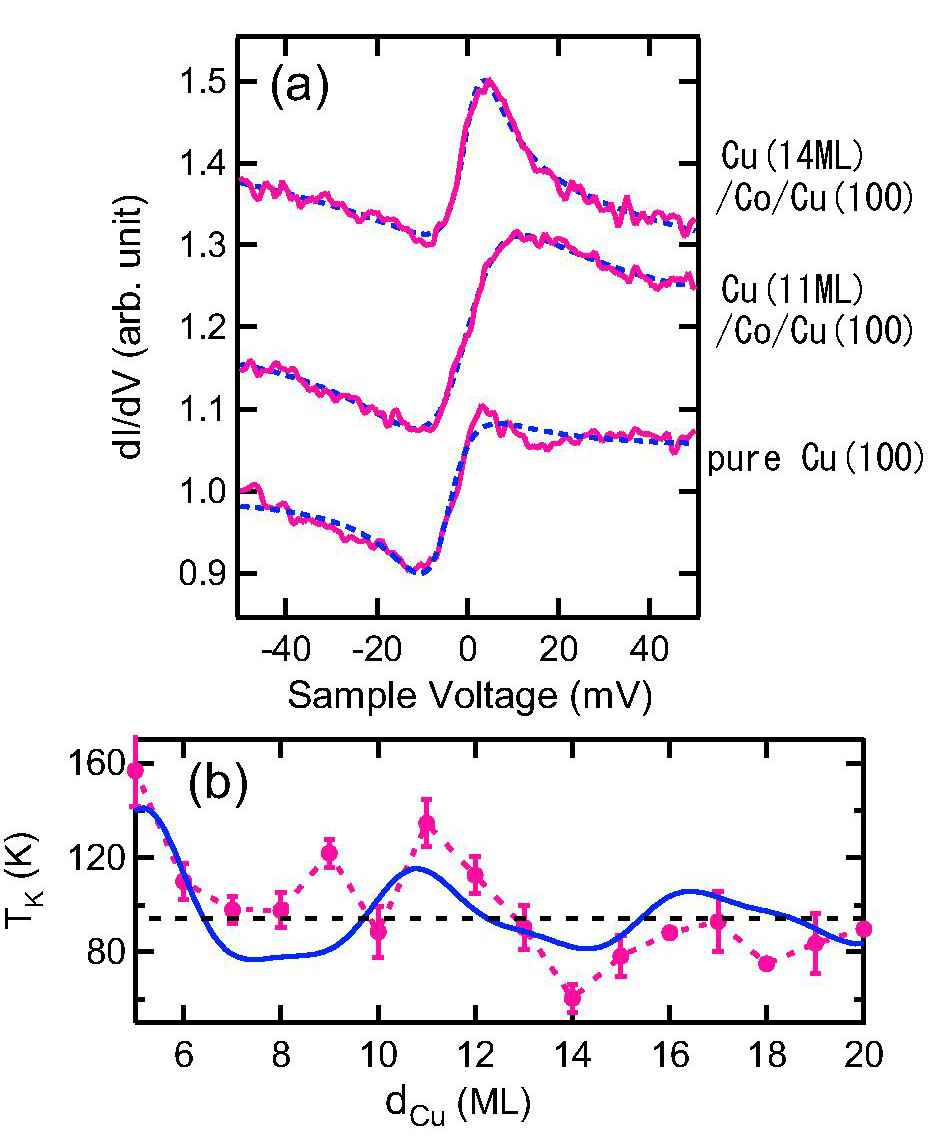}
  \caption{(Color online) (a) $\text{d}I/\text{d}V$ spectra measured on
  Co adatoms on Cu/Co/Cu(100) multilayers. Top: $14\,\text{ML}$ Cu. Middle:
  $11\,\text{ML}$ Cu. Bottom: pure Cu(100). Spectra are offset for clarity.
  Dashed lines indicate fits to experimental data according to
  Eq.\,(\ref{eq:FanoLine}). (b) Averaged Kondo temperature $T_{\text{K}}$ of
  Co adatoms on Cu/Co/Cu(100) multilayers as a function of Cu overlayer
  thickness $d_{\text{Cu}}$ (circles connected by dashed line). Half
  the error bar represents the standard error of $T_{\text{K}}$ calculated
  for each $d_{\text{Cu}}$. Calculated $T_{\text{K}}$ is plotted as a solid
  line. The fit parameters used are $A_{\text{b}}=0.046\,\text{eV}^{-1}\,\text{ML}$,
  $\Phi_{\text{b}}=0.26\pi$, $A_{\text{n}}=0.011\,\text{eV}^{-1}\,\text{ML}$,
  $\Phi_{\text{n}}=-0.03\pi$. Averaged $T_{\text{K}}=94\,\text{K}$ for Co
  adatoms on pure Cu(100) surfaces appears as a dotted line.}
  \label{Fig3}
\end{figure}
The top and middle graphs in Fig.\,\ref{Fig3}(a) are representative spectra for
$d_{\text{Cu}} = 14\,\text{ML}$ and $d_{\text{Cu}} = 11\,\text{ML}$, respectively
(solid lines). As a reference, a similar measurement on a pure Cu(100) substrate
is shown at the bottom of Fig.\,\ref{Fig3}(a). The characteristic line shapes asymmetric
around $V = 0$ indicate the Kondo resonance on these Co adatoms.
\cite{Neel_KondoContact,Wahl_KondoExchange}
In $\text{d}I/\text{d}V$ spectra the Kondo resonance generally appears as
an asymmetric line shape or even as a dip due to the Fano effect.
\cite{Ujsaghy_FanoTheory,Plihal_FanoTheory} This assignment was further
confirmed by disappearance of the characteristic line shape when the tip was
moved off the adatom center only by $0.5\,\text{nm}$.
\cite{Li_KondoCeAg,Madhavan_KondoCoAu}

To quantitatively analyze the $\text{d}I/\text{d}V$ spectra, we fit the
following Fano line shape to the data: \cite{Ujsaghy_FanoTheory,Plihal_FanoTheory}
\begin{equation}
  \frac{\text{d}I}{\text{d}V} =a\frac{(q+\epsilon)^2}{1+\epsilon^2}+b+cV
  \label{eq:FanoLine}
\end{equation}
with $\epsilon=(eV-\epsilon_{\text{K}})/k_{\text{B}}T_{\text{K}}$.
Here, $q$ is the asymmetry parameter of the Fano theory, $\epsilon_{\text{K}}$
a resonance shift from the Fermi level, and $a$ the amplitude of the resonance.
A linear voltage dependence $b+cV$ was added to account for the contribution
of conduction electron states not involved in the Kondo resonance.
Equation (\ref{eq:FanoLine}) was fitted to the above three spectra, with the
results shown in the same graphs as dashed lines. The Kondo temperatures
$67\,\text{K}$ and $139\,\text{K}$ were obtained for $d_{\text{Cu}} = 14\,\text{ML}$ and
$d_{\text{Cu}} = 11\,\text{ML}$, respectively, which significantly deviate
from $T_{\text{K}} = 90\,\text{K}$ for pure Cu (100). \cite{footnote_FanoFactor}

We performed analogous experiments for Cu/Co/Cu(100) multilayers with
Cu coverages ranging between $5$ and $20\,\text{ML}$.
The total number of samples and of spectra analyzed is 10 and 156, respectively. The Fano line
shape of Eq.\,(\ref{eq:FanoLine}) was fitted to all individual spectra to
obtain $T_{\text{K}}$ as described above. The whole data set was divided into
groups according to $d_{\text{Cu}}$, and the average and the standard error
of $T_{\text{K}}$ were calculated for each $d_{\text{Cu}}$. Figure \ref{Fig3}(b)
displays the averaged $T_{\text{K}}$ as a function of $d_{\text{Cu}}$ (circles connected by dashed line) with half the error bar representing the standard error. Measurements
on Co adatoms on pure Cu(100) were also repeated to obtain an averaged
$T_{\text{K}} = (94\pm 5)\,\text{K}\equiv T_{\text{K,0}}$, which is consistent
with previous reports. \cite{Neel_KondoContact,Wahl_KondoExchange}
$T_{\text{K,0}}$ is shown as a horizontal dashed line in Fig.\,\ref{Fig3}(b) as a
reference.

Clearly, the Kondo temperature $T_{\text{K}}$ is modulated as a function of
$d_{\text{Cu}}$.
This $T_{\text{K}}$ modulation is attributed to
variations in the density of states at the Fermi level, $\rho_{\text{F}}$,
caused by QW states of the Cu overlayer \cite{Xue_KondoPbQW}.
The Kondo temperature depends on $\rho_{\text{F}}$ through the following
relation: \cite{HewsonYoshida_Kondo}
\begin{equation}
  T_{\text{K}} = T_{0}\exp\left( -\frac{1}{2\rho_{\text{F}}J_{\text{sd}}} \right),
  \label{eq:TK_DOS_relation}
\end{equation}
where $T_{0}$ is a prefactor and $J_{\text{sd}}$ the $s$-$d$ exchange coupling
constant. This interpretation is plausible since we have proved the presence
of QW states and their evolution with $d_{\text{Cu}}$ in our samples. For a
Co adatom on pure Cu(100), inserting $\rho_{\text{F}} = 0.11\,\text{eV}^{-1}$,
\cite{footnote_CuDOS} $J_{\text{sd}}\approx 1.0\,\text{eV}$,
\cite{Ujsaghy_FanoTheory,Wahl_KondoSurfaces,Wahl_KondoExchange} and
$T_{\text{K,0}} = 94\,\text{K}$ into Eq.\,(\ref{eq:TK_DOS_relation}) gives
$T_{0}\approx 8900\,\text{K}$. $T_{\text{K}}$ is sensitive to a small variation
in $\rho_{\text{F}}$ under this condition. Since the quantum confinement in
the Cu/Co/Cu(100) multilayer is weak, $\rho_{\text{F}}$ at the surface of the
Cu overlayer is expressed by \cite{Qiu_PESQW,Bruno_ExchangeTheoryQW}
\begin{equation}
  \rho_{\text{F}}=\rho_{\text{F,0}} + \sum_{i=\text{b, n}} \frac{A_i}{d_{\text{Cu}}}\cos\left(\frac{2\pi d_{\text{Cu}}}{\Lambda_i} + \Phi_i \right).
  \label{eq:DOS_oscillation}
\end{equation}
Here, $A_i$, $\Lambda_i$, and $\Phi_i$ are the amplitudes, spatial periodicities,
and phases of the quantum oscillations at the belly ($i=\text{b}$) and the neck ($i=\text{n}$) of
the Cu Fermi surface. 
While the periodicities have been precisely determined ($\Lambda_{\text{b}}=5.88\,\text{ML}$ and $\Lambda_{\text{n}}=2.67\,\text{ML}$) \cite{Qiu_PESQW,Bruno_ExchangeTheoryQW}, the amplitudes and the phases are less established. 
We therefore fit $T_{\text{K}}$ calculated with Eqs.(\ref{eq:TK_DOS_relation}) and (\ref{eq:DOS_oscillation}) to the experimental data by keeping $\Lambda_i$ fixed and leaving $A_i$ and $\Phi_i$ as free parameters.
The result is plotted in Fig. 3(b) (solid line) with $A_{\text{b}}=0.046\,\text{eV}^{-1}\,\text{ML}$,  $\Phi_{\text{b}}=0.26\pi$, $A_{\text{n}}=0.011\,\text{eV}^{-1}\,\text{ML}$, and
  $\Phi_{\text{n}}=-0.03\pi$.
The fit reproduces most of the aspects of the data; local maxima at $\approx 5$, $\approx 11$, and $\approx 17\,\text{ML}$ are described as well as local minima at $\approx 7$ -- $8$ and at $\approx 14\,\text{ML}$. 
Note that the phases determined here deviate from a theoretical prediction of $\Phi_{\text{b}}=0.12\pi$ and $\Phi_{\text{n}}=0.69\pi$ for Cu/Co/Cu(100) multilayers \cite{Qiu_PESQW} by $0.14\pi$ and $-0.72\pi$, respectively.
This may be attributed to additional phase shifts due to the very presence of the Co adatoms.
Assuming that electron wavefunctions penetrate from the Cu surface to the vacuum side by a Co atomic height, this will result in an additional phase shift of $2\pi/\Lambda_{\text{b}}=0.34\pi$ and  $2\pi/\Lambda_{\text{n}}=0.75\pi$ for $\Phi_{\text{b}}$ and $\Phi_{\text{n}}$, respectively.
We thus call for theoretical investigations on the effect of Co adatoms on the phase shift of the quantum oscillations.

Finally, we remark that the exchange interaction between a Co adatom and the Co layer through Cu, $J_{\text{ex}}$, is too weak to account for the observed variations in $T_{\text{K,0}}$. 
According to an \textit{ab initio} calculation by Brovko \textit{et al.} \cite{Brovko_ExchangeTheory}, $J_{\text{ex}}$ is 1.5 meV for $d_{\text{Cu}}=5\,\text{ML}$ and rapidly decreases with increasing $d_{\text{Cu}}$.
This is small enough considering the energy scale of the Kondo effect $\text{k}_{\text{B}}T_{\text{K,0}} = 8.1\,\text{meV}$ observed here.
However, $J_{\text{ex}}$ was found to become comparable to or even much larger than $\text{k}_{\text{B}}T_{\text{K,0}}$ for $d_{\text{Cu}}\le 4\,\text{ML}$.
This should result in suppression and/or splitting of the Kondo resonance. 
\cite{Ralph_KondoMagnetic,Martinek_KondoMagnetic}
Furthermore, they predicted that the exchange interaction between two Co atoms adsorbed on a Cu/Co/Cu(100) multilayer can be tailored by changing the Cu overlayer thickness. \cite{Brovko_ExchangeTheory}
Our demonstration of tuning the Kondo effect by means of the same multilayer system promises experimental feasibility of such forthcoming studies.

In summary, we observed a modulation of the Kondo temperature of single
Co atoms adsorbed on Cu/Co/Cu(100) multilayers depending on the Cu layer
thickness. A model based on local density of states oscillations at
the Fermi level owing to confined QW states reproduces most of the
observed variations. 
The analysis on the $T_{\text{K}}$ modulation suggests unexpected phases of the quantum oscillations in the 
Cu overlayer, which requires further experimental and theoretical efforts.

Financial support by the Deutsche Forschungsgemeinschaft through SFB 668 is
acknowledged.


\begin{references}

\bibitem{HewsonYoshida_Kondo} A. C. Hewson,
\textit{The Kondo Problem to Heavy Fermions}
(Cambridge University Press, Cambridge, England, 1993);
K. Yoshida,
\textit{Theory of Magnetism} (Springer, New York, 1996).

\bibitem{Li_KondoCeAg} J. Li, W.-D. Schneider, R. Berndt, and B. Delley,
\prl \textbf{80}, 2893 (1998).

\bibitem{Madhavan_KondoCoAu} V. Madhavan, W. Chen, T. Jamneala, M. F. Crommie,
and N. S. Wingreen,
Science \textbf{280}, 567 (1998).

\bibitem{Manoharan_QuantumMirage} H. C. Manoharan, C. P. Lutz, and D. M. Eigler,
Nature \textbf{403}, 512 (2000).

\bibitem{Wiel_KondoQD} W. G. van der Wiel, S. De Franceschi, T. Fujisawa,
J. M. Elzerman, S. Tarucha, and L. P. Kouwenhoven,
Science \textbf{289}, 2150 (2000).

\bibitem{Nygard_KondoCNT} J. Nyg\r{a}rd, D. H. Cobden, and P. E. Lindelof,
Nature \textbf{408}, 342 (2000).

\bibitem{ParkLiang_KondoMolecule} J. Park, A. N. Pasupathy, J. I. Goldsmith, C. Chang, Y. Yaish, J. R. Petta, M. Rinkoski, J. P. Sethna, H. D. Abrun, P. L. McEuen, and D. C. Ralph,
Nature \textbf{417}, 722 (2002);
W. Liang, M. P. Shores, M. Bockrath, J. R. Long, and H. Park,
Nature \textbf{417}, 725 (2002).

\bibitem{Chen_KondoDimer} W. Chen, T. Jamneala, V. Madhavan, and M. F. Crommie,
\prb \textbf{60}, R8529 (1999).

\bibitem{Kliewer_Newns} J. Kliewer, R. Berndt, and S. Crampin, \prl \textbf{85}, 4936 (2000).

\bibitem{Limot_Step} L. Limot and  R. Berndt, Appl.\ Surf.\ Sci.\ {\textbf 237}, 576 (2004).

\bibitem{Neel_KondoContact} N. N\'eel, J. Kr\"oger, L. Limot, K. Palotas,
W. A. Hofer, and R. Berndt,
\prl \textbf{98}, 016801 (2007).

\bibitem{Hou_KondoMolecule} A. Zhao, Q. Li, L. Chen, H. Xiang, W. Wang, S. Pan, B. Wang, X. Xiao, J. Yang, J. G. Hou, and Q. Zhu,
Science \textbf{309}, 1542 (2005).

\bibitem{Iancu_KondoMolecule} V. Iancu, A. Deshpande, and S.-W. Hla,
\prl \textbf{97}, 266603 (2006).

\bibitem{Wahl_KondoExchange}P. Wahl, P. Simon, L. Diekh\"oner, V. S. Stepanyuk, P. Bruno, M. A. Schneider, and K. Kern,
\prl \textbf{98}, 056601 (2007).

\bibitem{Xue_KondoPbQW} Y.-S. Fu, S.-H. Ji, X. Chen, X.-C. Ma, R. Wu, C.-C. Wang, W.-H. Duan, X.-H. Qiu, B. Sun, P. Zhang, J.-F. Jia, and Q.-K. Xue,
\prl \textbf{99}, 256601 (2007).

\bibitem{Meier_ExchangeSPSTM} F. Meier, L. Zhou, J. Wiebe, and R. Wiesendanger, Science \textbf{320}, 82 (2008).

\bibitem{Brovko_ExchangeTheory} O. O. Brovko, P. A. Ignatiev, V. S. Stepanyuk, and P. Bruno, arXiv:0804.3298. 


\bibitem{CoSegregation} T. Bernhard, R. Pfandzelter, M. Gruyters, and H. Winter
Surf.\ Sci.\ \textbf{575}, 154 (2005).

\bibitem{Qiu_PESQW} Z. Q. Qiu and N. V. Smith,
J.\ Phys.: Condens.\ Matter \textbf{14}, 169 (2002).

\bibitem{Ortega_PESQW} J. E. Ortega and F. J. Himpsel,
\prl \textbf{69}, 844 (1992);
J. E. Ortega, F. J. Himpsel, G. J. Mankey, and R. F. Willis,
\prb \textbf{47}, 1540 (1993).

\bibitem{Bruno_ExchangeTheoryQW} P. Bruno,
\prb \textbf{52}, 411 (1995);
P. Bruno,
J.\ Phys.: Condens.\ Matter \textbf{11} 9403 (1999).

\bibitem{Crampin_CoCuQW} P. van Gelderen, S. Crampin, and J. E. Inglesfield,
\prb \textbf{53}, 9115 (1996).

\bibitem{Kawakami_PESQW} R. K. Kawakami, E. Rotenberg, E. J. Escorcia-Aparicio, H. J. Choi, T. R. Cummins, J. G. Tobin, N. V. Smith, and Z. Q. Qiu,
\prl \textbf{80}, 1754 (1998);
R. K. Kawakami, E. Rotenberg, E. J. Escorcia-Aparicio, H. J. Choi, J. H. Wolfe, N. V. Smith, and Z. Q. Qiu,
\prl \textbf{82}, 4098 (1999).

\bibitem{footnote_QWpeaks} Peaks rather than steps are expected because
only electronic states with small momenta parallel to the interface plane
effectively contribute to tunneling current [see R. C. Jaklevic, J. Lambe, M. Mikkor, and W. C. Vassell, \prl \textbf{26}, 88 (1971)].

\bibitem{Ujsaghy_FanoTheory} O. \'Ujs\'aghy, J. Kroha, L. Szunyogh, and A. Zawadowski,
\prl \textbf{85}, 2557 (2000).

\bibitem{Plihal_FanoTheory} M. Plihal and J. W. Gadzuk,
\prb \textbf{63}, 085404 (2001).

\bibitem{footnote_FanoFactor} We find that the Fano factor $q$ varies
significantly on Cu/Co/Cu(100) multilayers, reflecting changes in the spectrum
shape. This may be caused by quantum interference in the Cu overlayer.
\cite{Plihal_FanoTheory}

\bibitem{footnote_CuDOS} $\rho_{\text{F}}$ was estimated within the free
electron model.

\bibitem{Wahl_KondoSurfaces} P. Wahl, L. Diekh\"oner, M. A. Schneider, L. Vitali,
G. Wittich, and K. Kern, \prl \textbf{93}, 176603 (2004).

\bibitem{Ralph_KondoMagnetic} A. N. Pasupathy, R. C. Bialczak, J. Martinek, J. E. Grose, L. A. K. Donev, P. L. McEuen, D. C. Ralph, Science \textbf{306}, 86 (2004).

\bibitem{Martinek_KondoMagnetic} J. Martinek, M. Sindel, L. Borda, J. Barna\'s, J. K\"onig, G. Sch\"on, and J. von Delft, \prl \textbf{91}, 247202 (2003).

\end{references}
\end{document}